\documentclass[10pt]{article}

\usepackage{color}
\usepackage{hyperref}
\usepackage[flushleft]{threeparttable}
\usepackage{tabularx}
\usepackage{booktabs}
\usepackage{graphicx,subfigure}
\usepackage{amssymb}

\usepackage{natbib}

\usepackage[figuresright]{rotating}

\usepackage[margin=2cm]{geometry}

\begin{document}

\title{An Empirical Analysis of the Spatial Variability of Fuel Prices in the United States}

\author{Antonin Bergeaud$^{a}$ and Juste Raimbault$^{b,c,d,\ast}$\medskip\\
$^{a}$ Bank of France, Paris, France\\
$^{b}$ Center for Advanced Spatial Analysis, UCL, London, UK\\
$^{c}$ UPS CNRS 3611 ISC-PIF, Paris, France\\
$^{d}$ UMR CNRS 8504 G{\'e}ographie-cit{\'e}s, Paris, France\medskip\\
$^{\ast}$ Corresponding author: \texttt{juste.raimbault@polytechnique.edu}
}

\date{}

\maketitle

\begin{abstract}
In this paper, we use a newly constructed dataset to study the geographic distribution of fuel price across the US at a very high resolution. We study the influence of socio-economic variables through different and complementary statistical methods. We highlight an optimal spatial range roughly corresponding to stationarity scale, and significant influence of variables such as median income, wage with a non-simple spatial behavior that confirms the importance of geographical particularities. On the other hand, multi-level modeling reveals a strong influence of the state in the level of price but also of some local characteristics including population density. Through the combination of such methods, we unveil the superposition of a governance process with a local socio-economical spatial process. The influence of population density on prices is furthermore consistent with a minimal theoretical model of competition between gas stations, that we introduce and solve numerically. We discuss developments and applications, including the elaboration of locally parametrized car-regulation policies.\\

\end{abstract}

\section{Introduction}
\label{main}

\begin{figure}
\centering
\label{fig:1}
\subfigure[Time Series]{\label{fig:corr1}
	\includegraphics[width=0.4\linewidth]{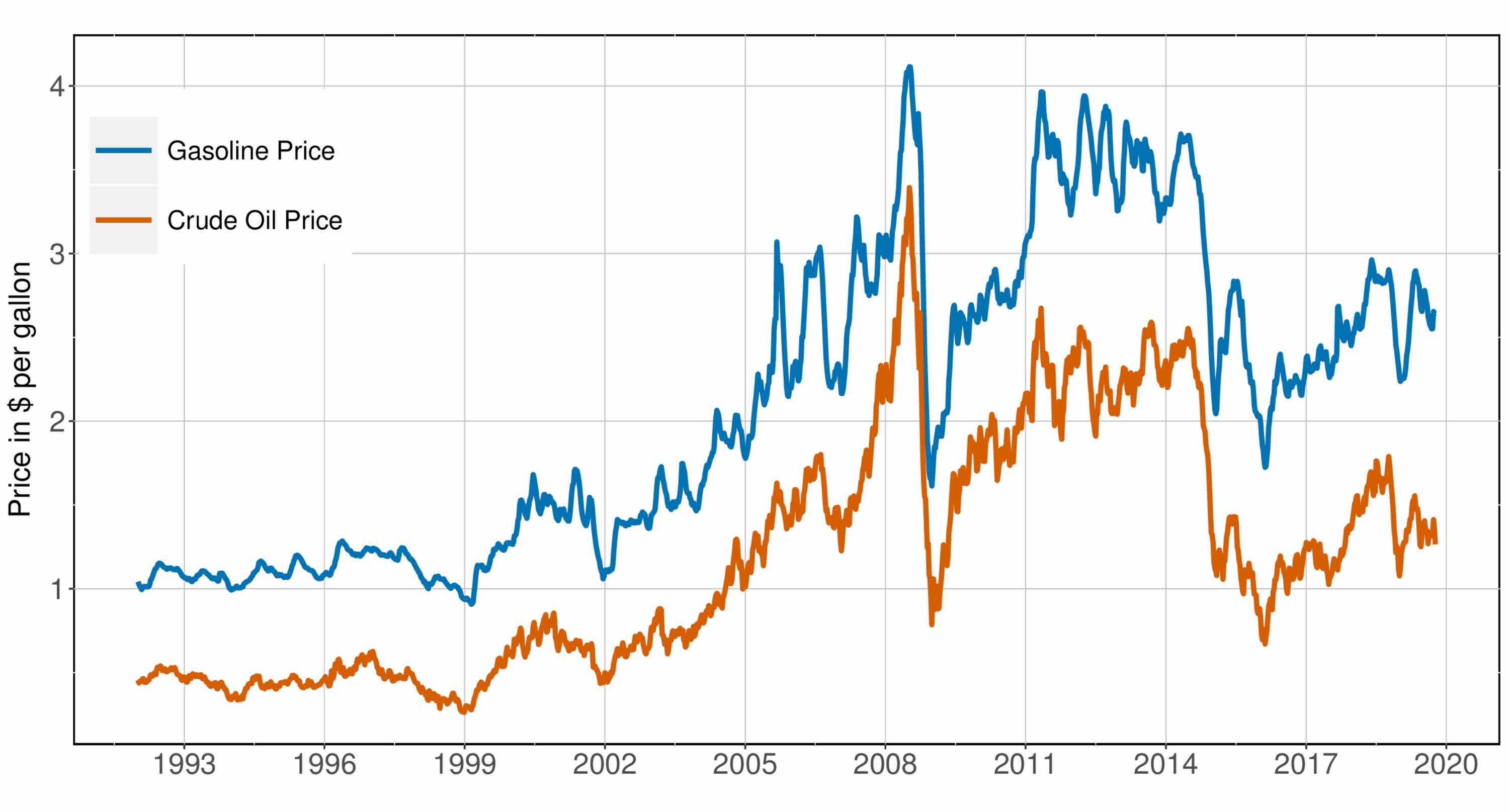}
    }
\subfigure[Correlation]{\label{fig:corr2}
	\includegraphics[width=0.4\linewidth]{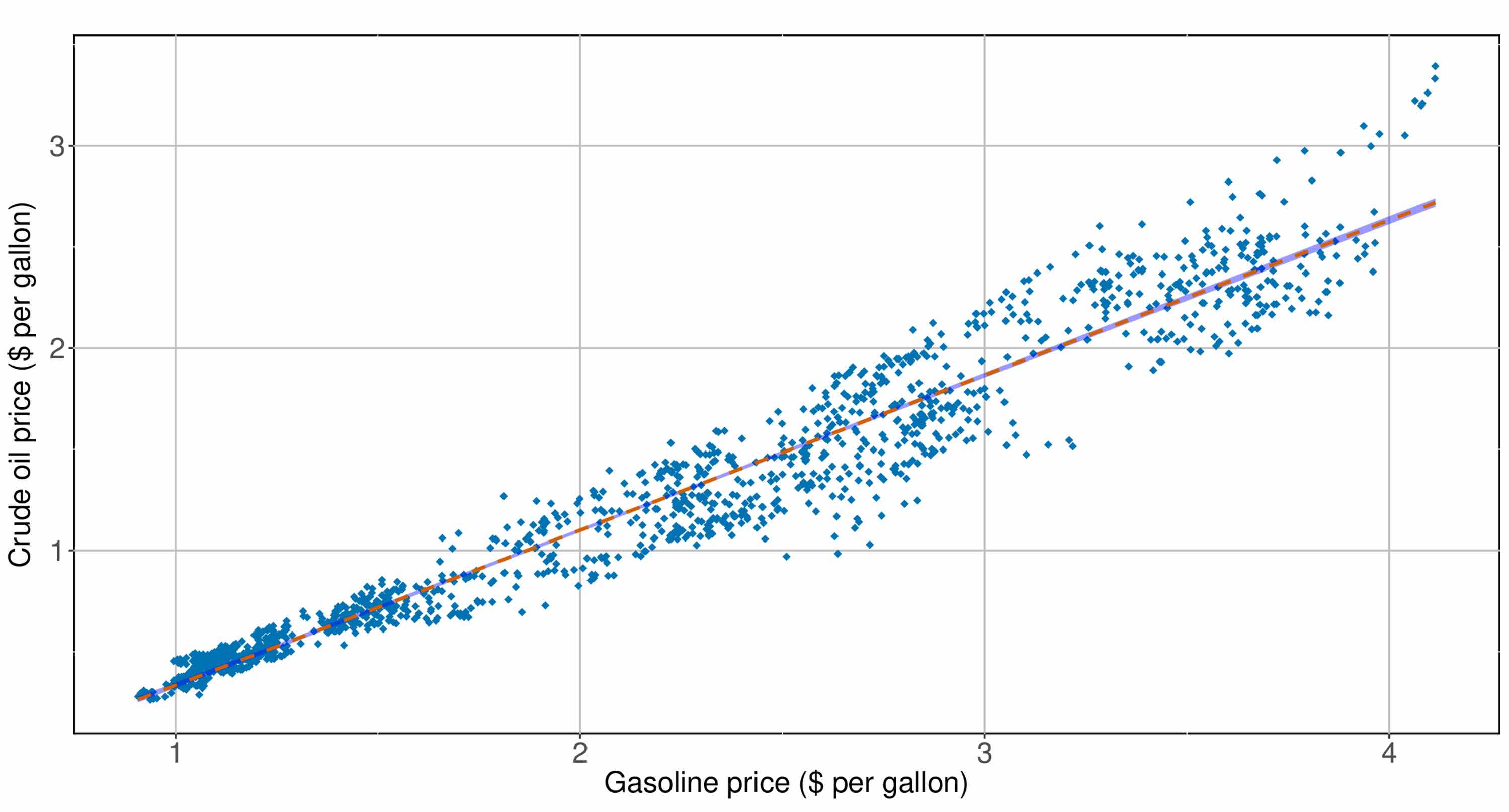}  
}
\caption{\textbf{Relationship between crude oil price and gasoline price (\$ per gallons)}. Left-hand side panel reports the weekly prices of both crude oil price of the West Texas Intermediate and average gasoline price in the US. Right-hand side panel reports the correlation and linear fit between the two. Linear model: y = -0.431[0.010]+0.766x[0.005], with $x$ the price of gasoline and $y$ the price of crude oil $R^2=0.949$, number of observations $N=1449$; standard errors in parentheses. Sources: EIA.}
\label{fig:corr}
\end{figure}

\begin{figure}
\centering
\label{fig:state}
\subfigure[Time Series]{\label{fig:ts_state}
	\includegraphics[width=0.4\linewidth]{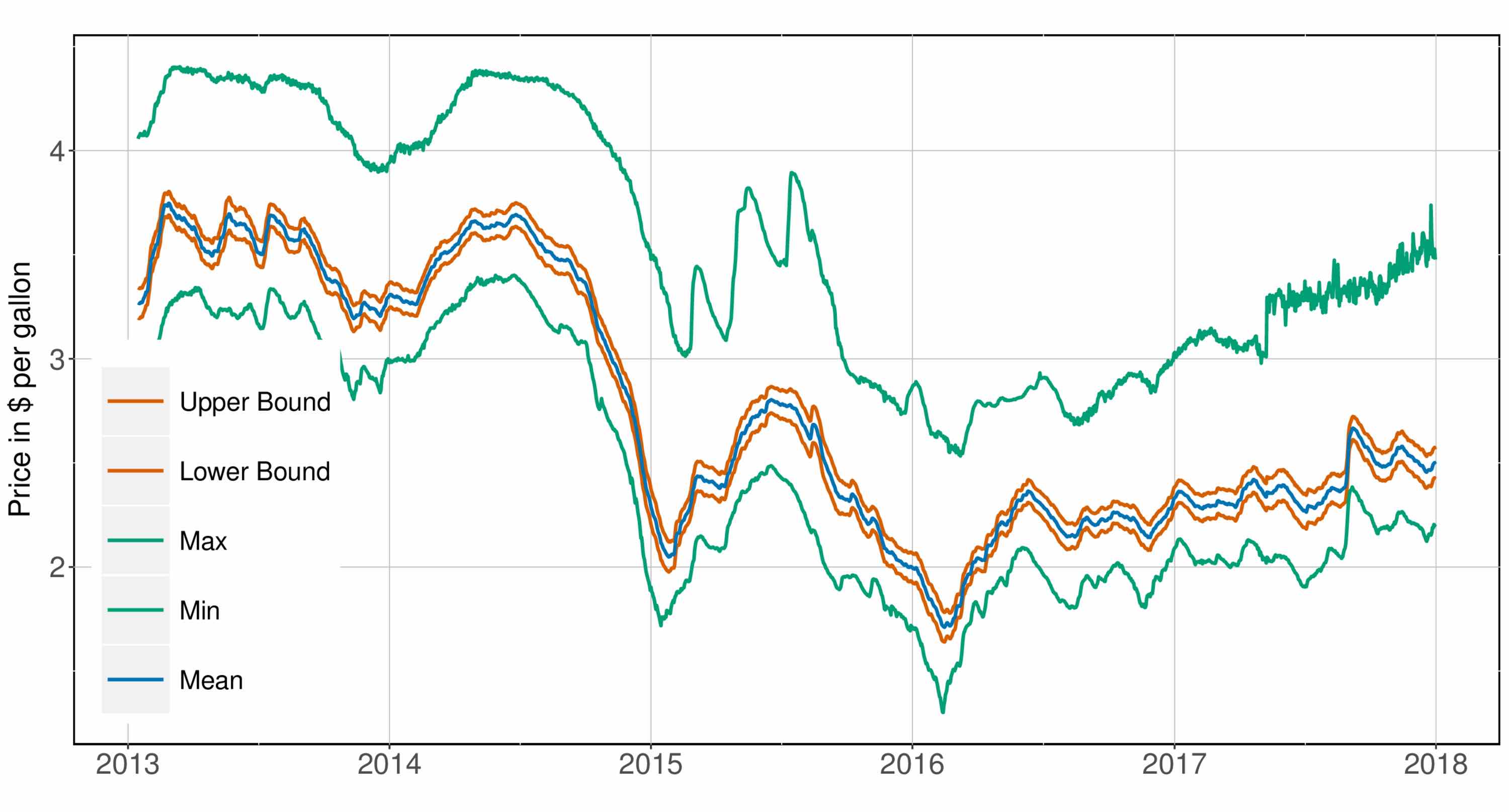}
    }
\subfigure[Coefficient of variation]{\label{fig:coeff_var_state}
	\includegraphics[width=0.4\linewidth]{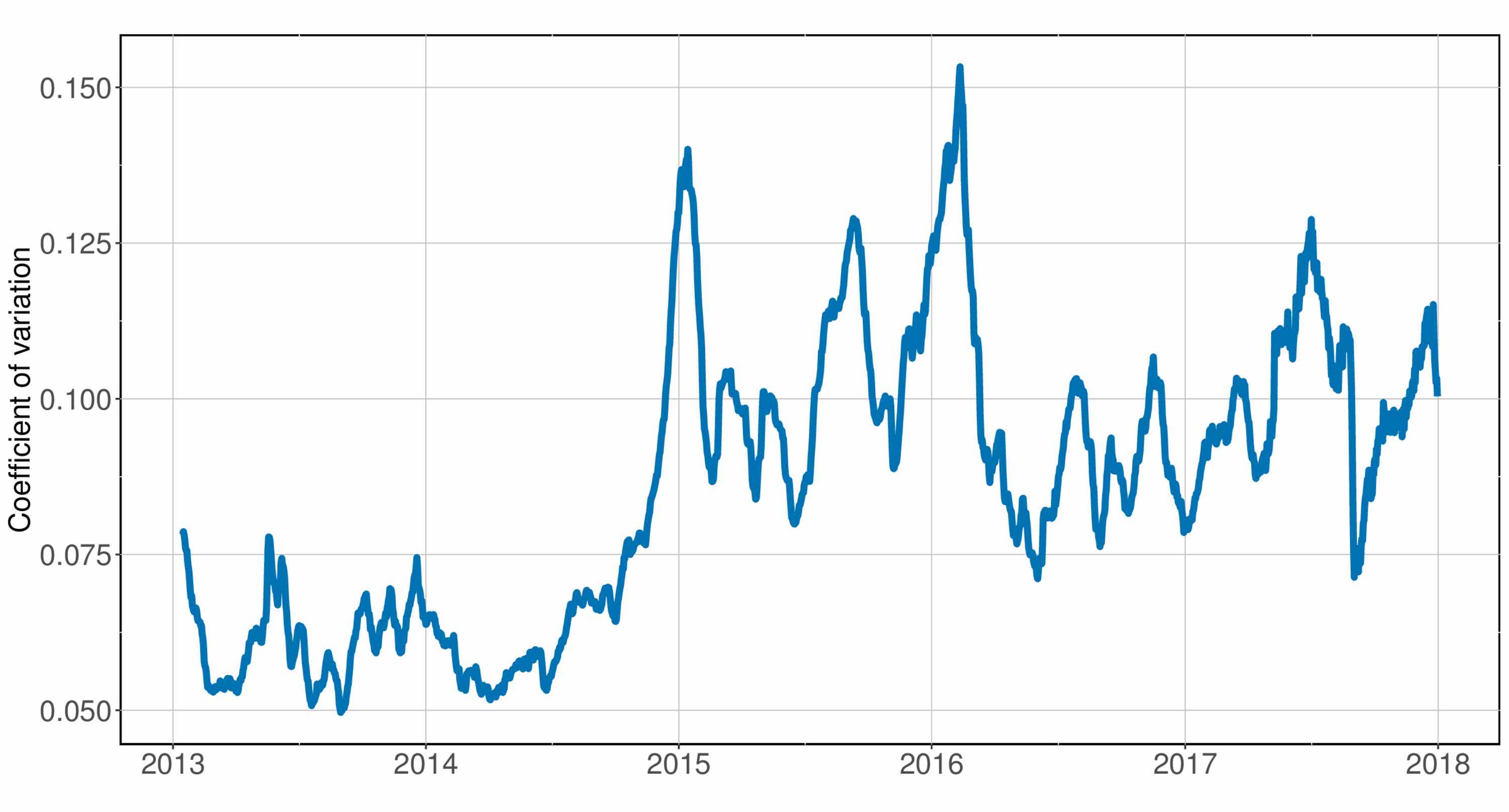}     
}
\caption{\textbf{Evolution and variation of average weekly prices (\$ per gallons)}. Left-hand side panel reports the unweighted average weekly price of gasoline across US states (excluding HI and AK), as well as plus and minus 2 standard errors intervals and maximum and minimum weekly values. Right-hand side panel reports the evolution of the weekly coefficient of variation (defined as the ratio of standard deviation over mean). Here again, HI and AK are excluded. Sources: EIA.}
\label{fig:ts}
\end{figure}

The distribution of the price of fuel is characterized by a large heterogeneity both in time and across geographical areas. Variations in time are relatively well understood and for a large part explained by changes in the price of crude oil \citep{borenstein1997gasoline}. This is confirmed by the Energy Information Administration (EIA) which states that a little more than half of the gasoline price is directly due to the underlying price of crude oil. It can also be easily confirmed by looking at the co-movements of these two prices, as we do in Figure \ref{fig:corr}. Surprisingly, although this correlation between crude oil price and gasoline is rather strong, significant differences can be observed across US states, as shown in Figure \ref{fig:ts}. 
Such spatial variations are of course mostly driven by state-specific gasoline tax rates (see e.g. \citealt{chouinard2004incidence}) that can be very different from one state to another. For example Pennsylvania adds a 58 cents while Texas only adds a 20 cents for each gallon of gasoline.\footnote{Figures are for January 2017 and do not include federal tax rate.} Yet, as we shall see, the spatial distribution of gasoline price also depends in other factors. Such exploratory spatial analysis is the object of the present paper.

\begin{figure}
\centering
\includegraphics[width=0.6\linewidth]{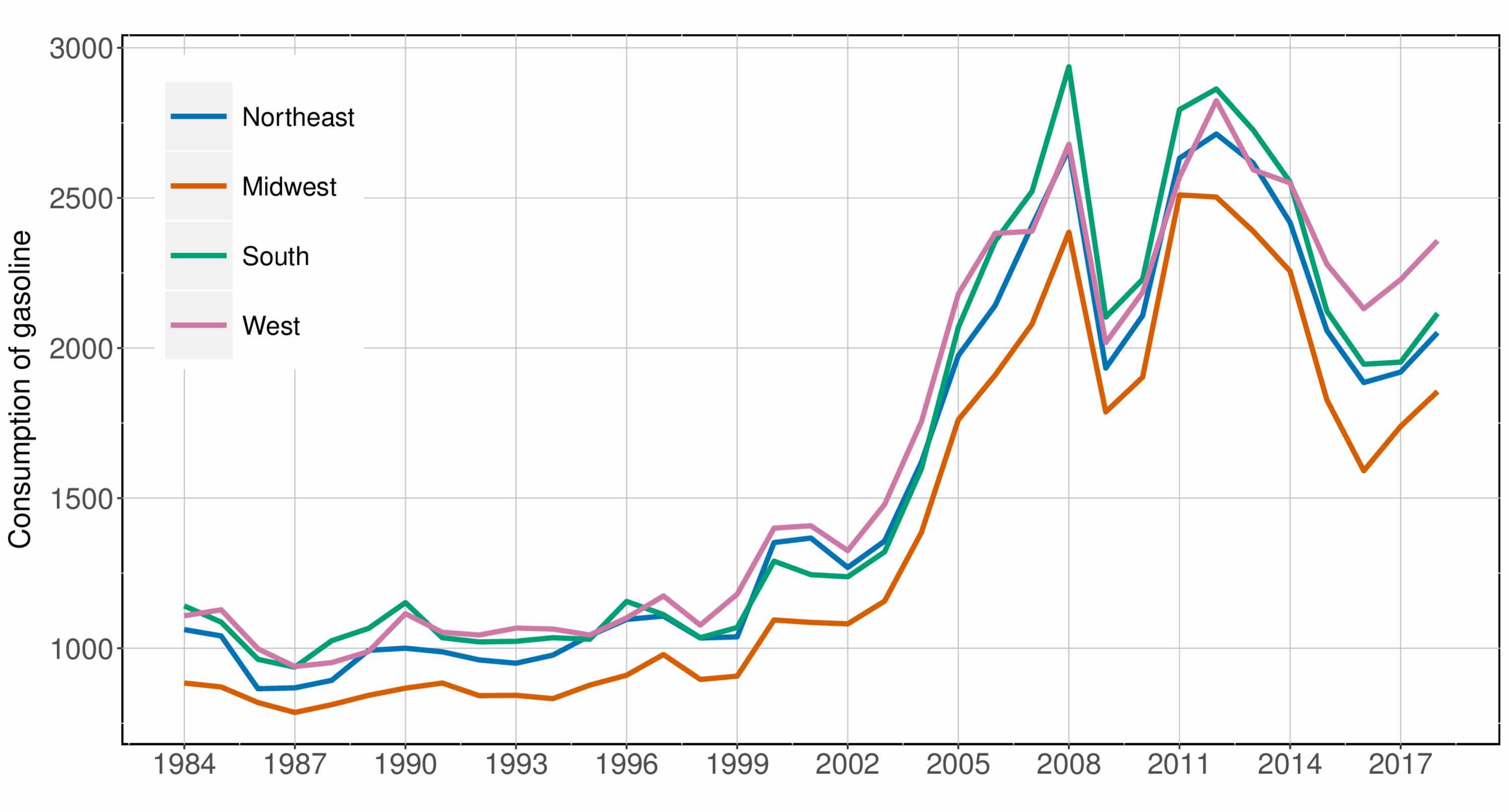}
\caption{\textbf{Consumption of gasoline and oil per household by region}. Source: Consumer Expenditure Survey, 1984-2016. In constant dollars per year \label{fig:consumption_gas}
}
\end{figure}

There exists to our knowledge no systematic mapping in space and time of retail fuel prices for a country as large as the US. The main reason is probably that the availability of data continue to be a significant obstacle to such research. It is also likely that the nature of the problem itself may pose some difficulties as it lies at the crossroad of several disciplines. While economists study price elasticity and measurement in different markets, transportation geographers apply methods such as transportation prices in spatial models to emphasize spatial patterns, more than they focus on precise market mechanisms. On the contrary, gas price is a very frequent subject of attention by politics, media and consumers and its variation is very closely scrutinized by financial analysts and transport professionals. The price of oil is also an important driver of the level of inflation and the cost of living, with a growing importance since, as reported in Figure \ref{fig:consumption_gas}, households' consumption of motor oil has been increasing rapidly in the past decades.

In this paper, we use a new database on gasoline prices at a very local level collected during two months from January to March 2017, to explore their variability across time and space, even within-states. This is a first exploration of what such microdata can teach us about the distribution of the retail price of gas and should not be taken as a complete analysis of the determinant and predictive drivers of the variation of the cost of fuel.\footnote{To carry out such analysis, one would probably require a larger time dimension in order to exploit exogenous variations in local socio-economic conditions to measure the subsequent impact on gasoline price. We discuss this at the end of this paper.} These variations can have many causes, from the crude oil price to local tax policy, infrastructure and geographical features, and even strategic interactions between competitors, all having heterogeneous effects in space and time. For example, oil spill events, such as the BP's Deepwater Horizon accident in the Gulf of Mexico in 2010, have a large impact on the prices at the pump, which is likely to be heterogeneous in space.\footnote{To our knowledge, whether or not the effect of the Deepwater Horizon oil spill affected gasoline price unequally is still under debate by experts. Typically, such microdata could help addressing this question. We however leave this analysis for further research.} Similarly, disruption in service of some specific pipelines could affect some regions but not others (see e.g. \citealt{McRae2017}). But local variations in prices can also reflect heterogeneity in more indirect socioeconomic indicators such as territorial inequalities, geographical singularities and consumer preferences.

Nevertheless, examples of somehow related works can be found in the literature. For example,~\cite{rietveld2001spatial} study the impact of cross-border differences in fuel price and the implications for gradual spatial taxation in Netherlands. At the country level, \cite{rietveld2005fuel} provide statistical models to explain fuel price variability across European countries based on fiscal competition. Such competition has some negative effects as it encourages consumers to cross the border in order to refill which make them drive more. Given that large variations in oil prices across some US-state borders can be important (according to Gasbuddy, a crowd-sourced information website aiming at ``\textit{nowcasting}'' gasoline price in North America, some borders like the AZ-CA state line can induce change in the price of filling a tank by up to \$25)\footnote{\href{https://business.gasbuddy.com/blog\-overpaying\-across-us-borders/}{See their blog article on this topic}}, such rational behaviors are clearly at play in the US as well. Relatedly, \cite{macharis2010decision} model the impact of spatial fuel price variation on patterns of inter-modality, implying that the spatial heterogeneity of fuel prices has, as we would expect, a strong impact on user behavior. With a similar view on the geography of transportation, \cite{gregg2009temporal} study spatial distribution of gas emission at the US-state level.

The geography of fuel prices also have important implications for public economists. Of course the economic implication of an increase in oil price can be quite dramatic as documented by \citet{kilian2008}; \citet{hamilton2009} and many others, given how central oil and transport are in the production network and global value chains. Yet, one major interest economists have found looking at the spatial dynamics of oil price, and the subsequent consumer reaction, is to learn how to design an effective tax system (see e.g. \citealt{parry2005does}) that correct for the negative externalities associated with the use of automobile (pollution, accidents, congestion, noise etc... see \citealt{parry2007automobile} for a review on these topics). \citet{li2014gasoline} conduct such an analysis by examining consumers' reaction to a shock in the price of gas induced by a tax increase (as opposed to a supply shock) and are able to quantify the elasticity of consumption to price. \citet{Hughes2006} show that the underlying preferences shaping such elasticity have changed a lot in time while \citet{bento2009} find that each cent-per-gallon increase in the price of gasoline reduces consumption by about 0.2 percent, with large heterogeneity across household revenue. In France, \cite{combes2005transport} accurately measure transportation costs across urban areas using the price of fuel as one major driver in order to derive regional policy recommendations.

A related question of interest by economists is to understand the dynamics of price adjustment. While frictions in adjusting costs of goods to supply and demand shocks have been widely studied by economists, theoretically and empirically, fuel prices offer a very nice case study given the frequency of stimuli on its cost. Thus, and closely related to our work in their use of daily open data of fuel price for France, \cite{gautier2015dynamics} investigate dynamics of transmission from crude oil prices to fuel retail prices. However, they do not introduce an explicit spatial model of prices diffusion and do not study spatio-temporal dynamics. They estimate the pass through of wholesale prices to retail gasoline prices to be equal to 0.67, meaning that on average, two-thirds of an increase in crude oil price translates to an increase in price at the pump. Similar analysis have been conducted in other countries, for example in Sweden by \citet{asplund2000price} or in Spain by \citet{stolper2016bears}. Similarly \cite{deltas2008retail} studies a long-time panel of prices in the US to show an asymmetry in response to wholesale price variations, but restricts the analysis at the State level. A review on empirical studies of gasoline retailing done by~\cite{eckert2013empirical} confirms that station-level studies are few and generally consider small countries such as Belgium or Austria, or single US States (for example \citealt{hosken2008retail} in DC).

\begin{table}
	\centering
\caption{Example of processes potentially involved in the formation of retail fuel prices, at different scales.}
\medskip
\label{tab:processes}
\begin{tabular}{|p{5cm}|p{5cm}|p{5cm}|}
\hline
\textbf{Local} & \textbf{Regional} & \textbf{National/global}  \\
\hline
Local competition, socio-economic variables, retailer choices, consumer choices, accessibility, availability of transportation alternatives & Regional taxes, geographical situation, fuel transportation prices, road infrastructures, oil transportation and refinement infrastructures, culture  & National taxes, crude oil prices, company strategies, geopolitical factors \\
\hline
\end{tabular}
\end{table}

In this paper we adopt a broader approach by proceeding to exploratory spatial analysis on fuel prices for the whole US. We show that most of the variation occurs between counties and not across time, although crude oil price was not constant during the period considered. Among the potential processes involved in the formation of retail fuel prices described in Table \ref{tab:processes}, a large majority of them are tightly linked to spatial structure or geography, which we propose to investigate as the main purpose of this paper. We therefore turn to a spatial analysis of the distribution of fuel prices. Our main findings are twofold: first we show that there are significant spatial pattern in some large US regions, second we show that even if most of the observed variation can be explained by state level policies, and especially the level of tax, some county specific characteristics still play a significant role in explaining the variance of prices.

The rest of the paper is organized as follows: in next section, we describe a generic procedure and the tool used for a systematic data collection. We also present our dataset. In section \ref{sec:result} we conduct spatial statistics analysis in order to study the variation of fuel price across time and space and test the potential correlation with some covariates. We then develop in section \ref{sec:econometry} an econometric analysis to study more precisely the role of these covariates, and develop a minimal theoretical model linking population density with prices which appears to be consistent with our econometric results. Finally, in section \ref{sec:discuss} we discuss our results and conclude.

\section{Dataset} \label{sec:data}

Our dataset contains daily information on fuel price at the gas station level for the whole US mainland territory. These information have been constructed from self-reported fuel price and span almost the entire universe of gas station in the US. We start by describing data collection and then give some statistics about this new dataset.

\subsection{Collecting large scale heterogeneous data}

The availability of new type of data has induced consequent changes in various disciplines from social science (e.g. online social network analysis, see \citealp{tan2013social}) to geography (e.g. new insights into urban mobility or perspectives on ``smarter'' cities, see \citealp{batty2013big}) including economics where the availability of exhaustive individual or firm level data is seen as a revolution of the field. Most studies involving these new data are at the interface of implied disciplines, which is both an advantage but also a source of difficulties. For example misunderstandings between physics and urban sciences described in \cite{dupuy2015sciences} are in particular caused by different attitudes towards unconventional data or divergent interpretations and ontologies. Transportation systems are naturally at the core of such problematic involving new ``big data'' production, collection and processing \citep{kitchin2014real}.

Collection and use of new data has therefore become a crucial stack in social-science. The construction of such datasets is however far from straightforward because of the incomplete and noisy nature of the underlying data. Specific technical tools have to be implemented but have often been designed to overcome one specific problem and are difficult to generalize.
 
We develop here such a tool that fills the following constraints that are typical of large scale data collection, namely a reasonable level of flexibility and generality, an optimized performance through parallel collection jobs, and the anonymity of collection jobs to avoid any possible bias in the behavior of the data source. The architecture, at a high level, has the following structure: (i) an independent pool of tasks runs continuously socket proxies to pipe requests through \texttt{tor}; (ii) a manager monitors current collection tasks, split collection between subtasks and launches new ones when necessary; (iii) subtasks can be any callable application taken as argument destination urls, they proceed to the crawling, parsing and storage of collected data. The application is open and its modules are reusable: source code is available on the repository of the project.\footnote{at \texttt{https://github.com/JusteRaimbault/EnergyPrice}} We constructed our dataset by using the tool continuously in time during two months to collect crowdsourced data available from various online sources.

\subsection{Dataset}

Our dataset comprises around $41\cdot 10^6$ unique observations of retail fuel prices at the station level, spanning the period starting the $10^{th}$ of January 2017 and ending the $19^{th}$ of March 2017 and corresponds to 118,573 unique retail stations. For each of these stations, we associate a precise geographical location (at the city level). On average we have 377 price information by station. Prices correspond to various possible fuel types but we focus the study to this ``regular'' fuel which corresponds to a third of the observation points and 4,629 points per county on average.\footnote{We let potential analyzes of the different dynamics across fuel type for further research.} A few outliers resulting of coding errors were also filtered from the dataset.

Our final dataset thus contains 14,192,352 observations from 117,155 gas stations, followed during 68 days. As the data was collected continuously, more than one information per day can be collected for some stations in which case we further collapse these data by day, and take the average of the observed price per gallon, to obtain a panel of 5,204,398 gas station - day observations.\footnote{The panel is not balanced as prices are not reported every day in some stations. The average gas station has information on price for 44 days (over 68).}  Table \ref{tab:stat_desc} gives some basic descriptive statistics of on price data showing that the distribution of oil price is highly concentrated with a small skewness (the ratio of the $99^{th}$ to the $1^{st}$ percentile is 1.6).


Regarding the spatial distribution of stations themselves, at the level of zipcodes the number of stations is naturally correlated with population with a Pearson correlation of 0.72. However, when considering zip areas with a population larger than the median (resp. the third quartile), including 90\% (resp. 65\%) of the total population, this correlation drops to 0.52 (resp. 0.29). This means that in urban areas, which broadly correspond to the most populated zip codes, the local density of gas stations is loosely correlated with population\footnote{Note that this result implies that scaling relationships identified at the city level, for example by \cite{bettencourt2007growth}, are not robust to a change in spatial resolution. Further investigations of these properties remain however out of the scope of this paper.}. This legitimates in a first approximation the assumption of exogenous location of stations we will do in our theoretical model below (section~\ref{sec:theory}): the low correlation is considered as being zero and locations considered as random regarding population density.

Finally, in the spatial analysis, we will also use socioeconomic data at the county level, available from the various administrative sources. We will use the latest available, which can imply relying to the 2010 Census. In the first preliminary analysis, we use basic socio-economic variables (income, population, average wage, jobs per capita, jobs), while more elaborate covariates are added thereafter in the econometric analysis (unemployment, poverty, political preferences etc...). All variables are aggregated at the county level. Note that the goal of this exercise is not to give a structural estimation of the determinant of gas price. These variables are proxies that aim at capturing correlations. The actual mechanism through which they affect the dynamic of oil price is not modelled here. We however provide an example of how a theoretical model can highlight such mechanism in section \ref{sec:theory}. Details and source for these county level variables are given in Table \ref{tab:county_var_desc}.

\begin{table}
\centering
\caption{Descriptive statistics on Fuel Price (\$ per gallon)}
\medskip
\label{tab:stat_desc}
\begin{tabular}{ccccccc}
\textbf{Mean} & \textbf{Std. Dev.} & \textbf{p10} & \textbf{p25} & \textbf{p50} & \textbf{p75} & \textbf{p90} \\
\hline
\cr
2.28 & 0.27 &  2.02  &  2.09  &  2.21  &  2.39  &  2.65  \\
\hline
\end{tabular}
\end{table}

\section{Exploratory analysis} \label{sec:result}

\subsection{Spatio-temporal Patterns of Prices}\label{subsec:patterns}

\begin{figure}
\centering
\subfigure{\label{fig:map_price}
	\includegraphics[width=\linewidth]{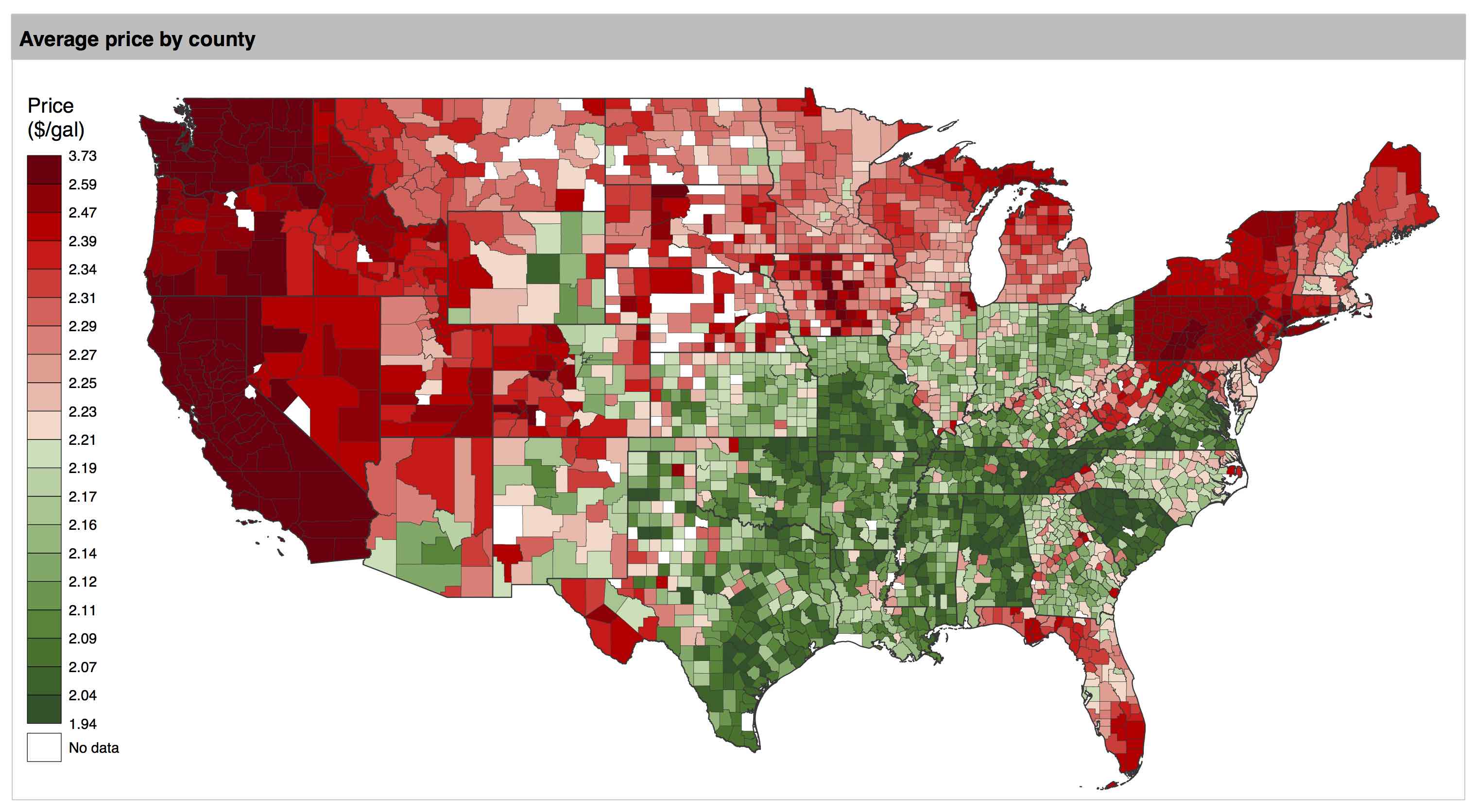}
    }\\
\subfigure{\label{fig:map_price_notaxes}
	\includegraphics[width=\linewidth]{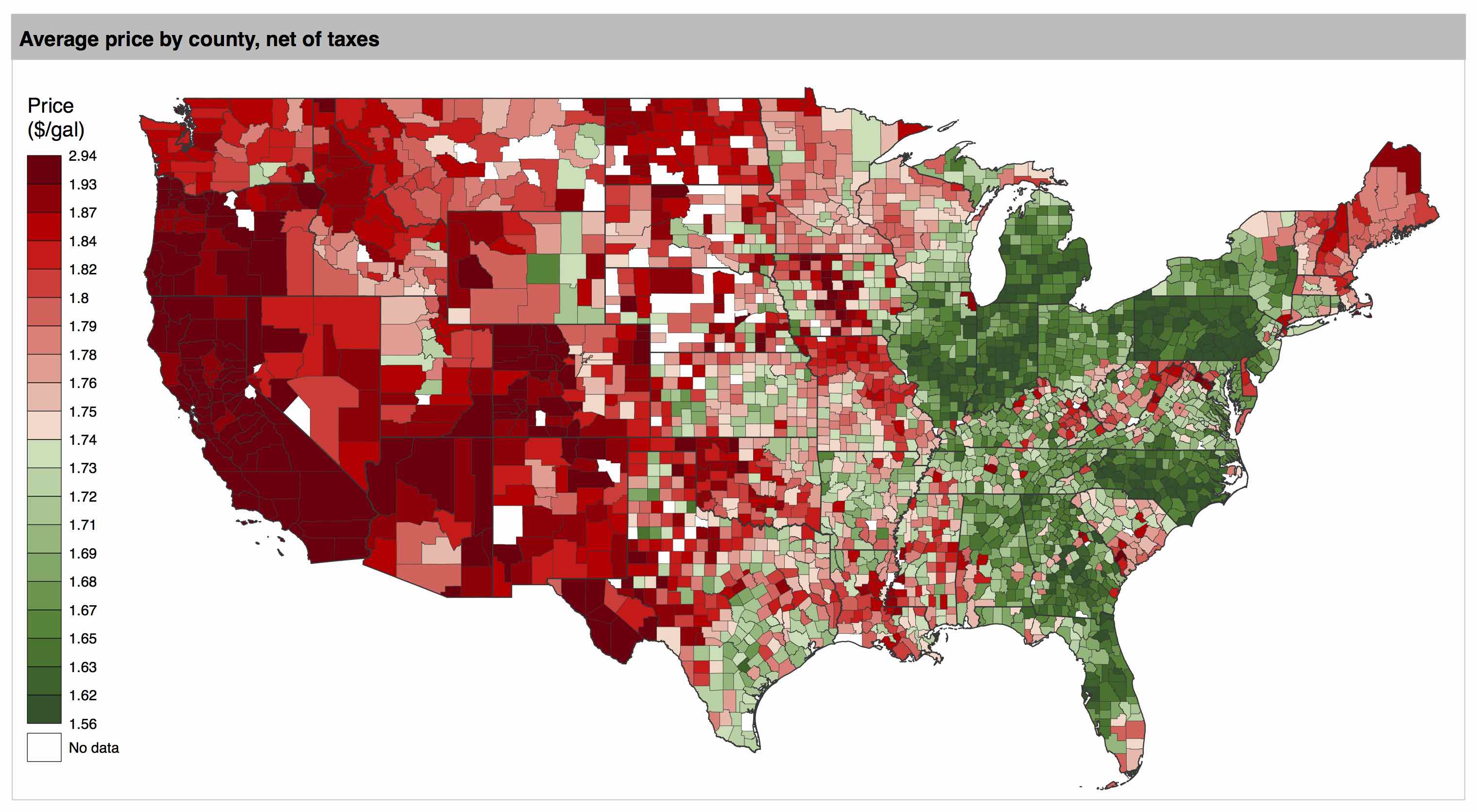}     
}
\caption{(Top) Map of mean price for counties, regular fuel, averaged over the whole period. (Bottom) Same map with prices net of taxes.}\vspace{-0.3cm}
\label{fig:maps_price}
\end{figure}

Before moving to a more systematic study of the variation of fuel price, we propose a first exploratory introduction to give insight about its structure across time and space. This exercise is a crucial stage to guide further analyses, but also to understand their implications in a geographical context. To explore the data, we built a simple web application which allow to map gas price in space and time. This application is available  \href{https://analytics.huma-num.fr/geographie-cites/fuelprice}{online} and allow the user to browse for our data easily. We also show one example of mapping the data at the county level in Figure \ref{fig:map_price} where we have aggregated the data over the whole period at the county level using an unweighted mean over all observations. This map display quite clear regional patterns with the South-central and Southeast regions having the lowest prices and the Pacific cost and Northeast the highest prices. Some geographical structures and discrepancies in underlying processes of price formation can already be suggested when comparing the price map with the same map when taxes have been removed (Figure~\ref{fig:map_price_notaxes}). The West coast still exhibits the higher prices, although it has high level of taxes, whereas in some States such as Pennsylvania, State of New York, or Florida high prices are essentially due to taxes. On the contrary, taxes have close to no effect in South Carolina. This suggest an interplay between geographical and political processes, which will be explored in more details below.

Of course, plotting aggregated data over the whole period does not bring much information about the time variation of the data. In fact, most of the variation of fuel price occurs across space, at least during the relative short period of observation we are considering. A variance decomposition of fuel price shows that only 11\% of the total variance is explained by \emph{within} gas station variations. Similarly, the Spearman's rank correlation coefficient between the gas station price of regular fuel in the first day of dataset and in the last day is 0.867, and the null hypothesis that these two information are independent is strongly rejected.

\begin{figure}
\centering
\label{fig:moran}
\subfigure[Time Series]{\label{fig:moran_day}
	\includegraphics[width=0.4\linewidth]{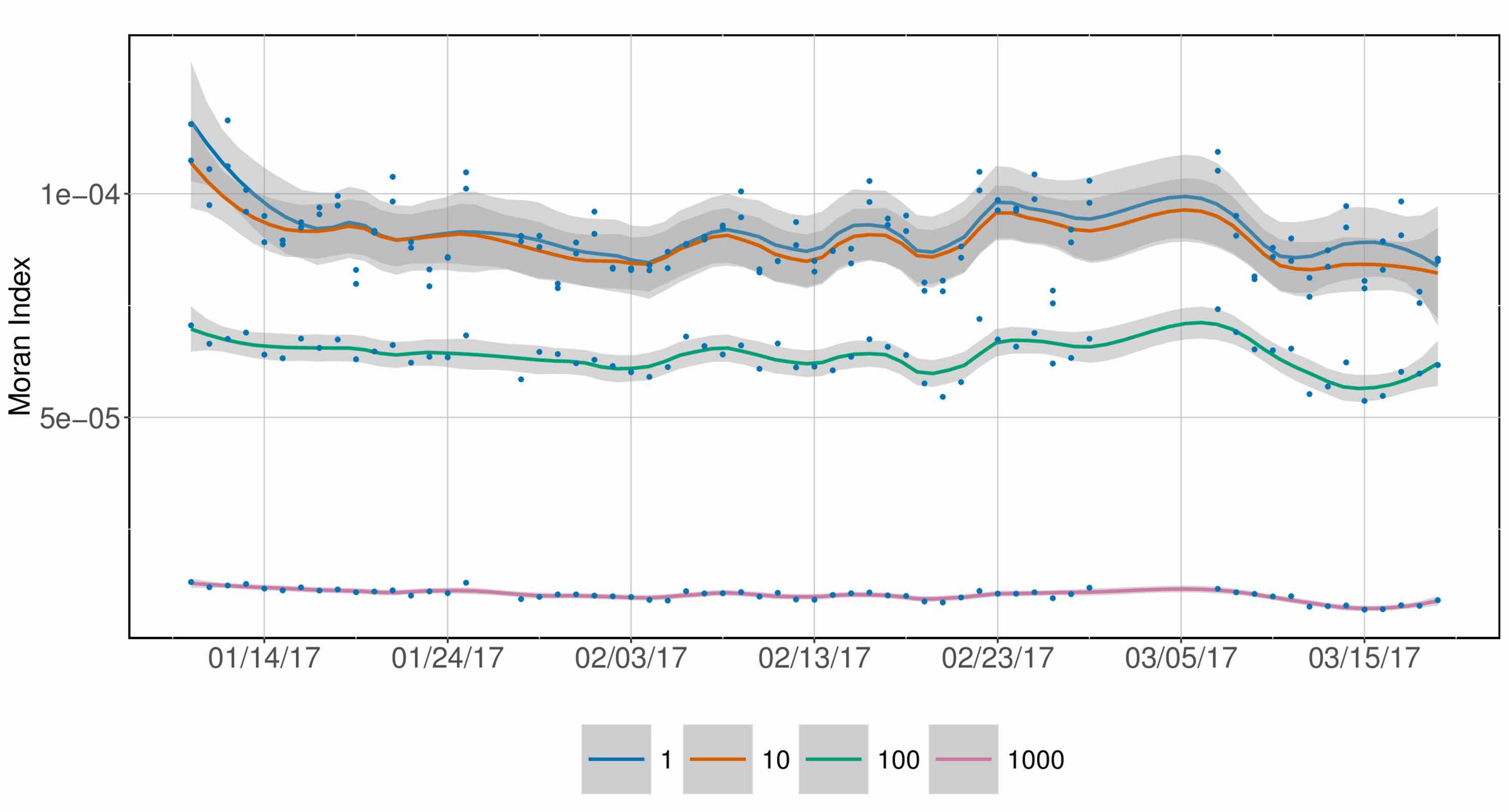}
    }
\subfigure[Decay]{\label{fig:moran_week}
	\includegraphics[width=0.4\linewidth]{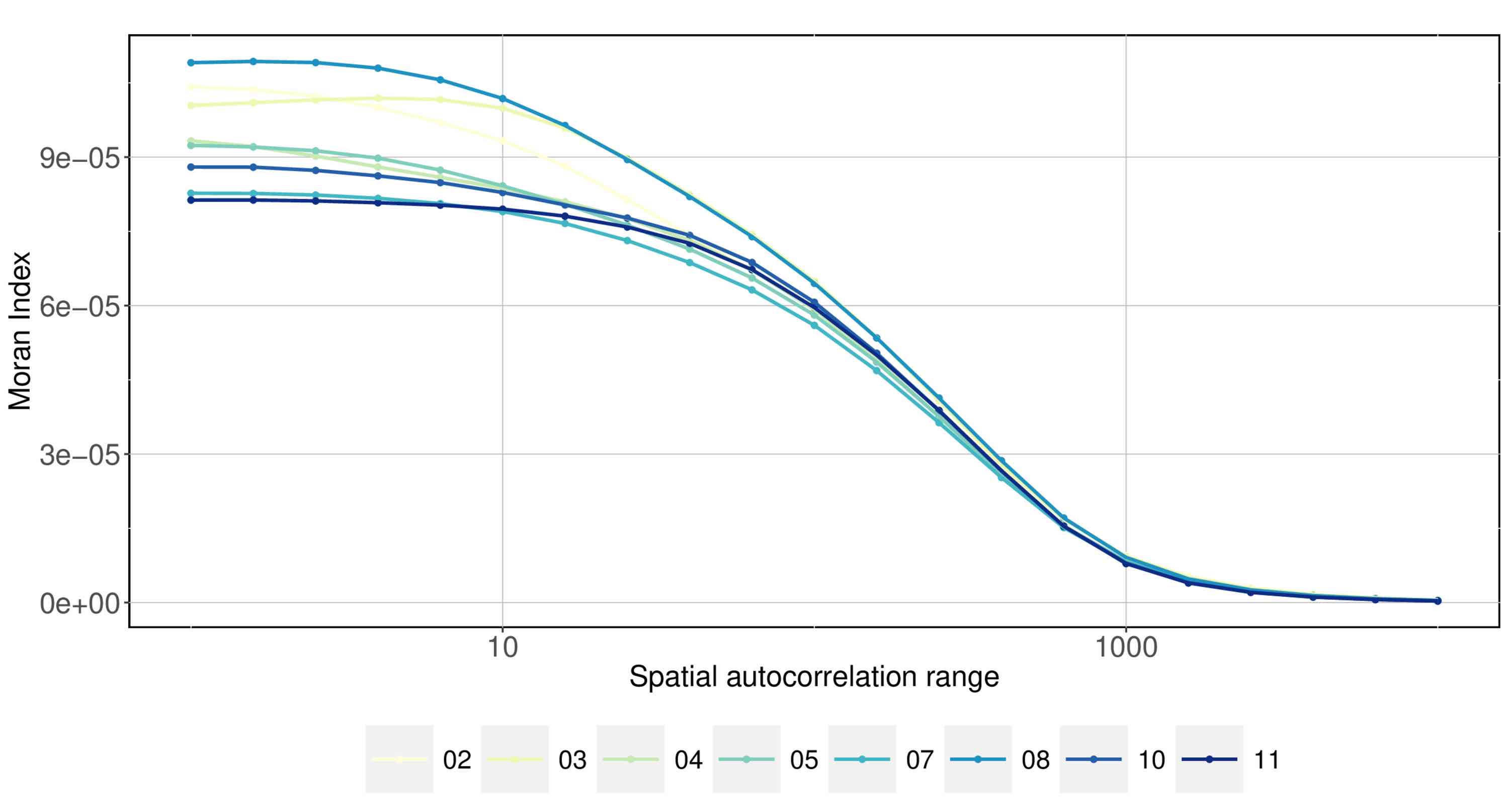}     
}
\caption{\textbf{Behavior of Moran spatial-autocorrelation index.} (Left) Evolution in time of Moran index computed on daily time windows, for different decay parameter values. (Right) Moran index as a function of decay parameter, computed on weekly time windows.}
\end{figure}

Since most of the variation in oil price is \emph{between} gas station, we now focus mainly on spatial correlations. We will conduct the analysis at the county level for various reasons. First it appears that a variance decomposition of fuel price between and within county shows that more than 85\% of the variance is between-county, second because the localization of gas station is not reliable enough to allow for a smaller granularity and third because we have many socioeconomic information at this level. We therefore study the spatial autocorrelation of prices at the county level. Spatial autocorrelation can be seen as an indicator of spatial heterogeneity which we measure using the Moran index as in \cite{tsai2005quantifying}, with spatial weights of the form $\exp{\left(-d_{ij} / d_0 \right)}$ where $d_{ij}$ is the distance between spatial entities $i$ and $j$, and $d_0$ a decay parameter that captures the spatial range of interactions accounted for in the computation. We show in Fig.~\ref{fig:moran} its variations for each day and also as a function of the decay parameter. Intuitively, this parameter capture how the spatial correlation of price change with the distance between stations.

The fluctuations in time of the daily Moran index for low and medium spatial range, confirms geographical specificities in the sense of locally changing correlation regimes. These are logically smoothed for long ranges, as price correlations drop down with distance. The behavior of spatial autocorrelation with spatial range is particularly interesting: we observe a first change of regime at around 10km (from constant to piece-wise linear regime), and a second important one at around 1000km, both consistent across weekly time windows (see Figure \ref{fig:moran_week}). We conjecture that these correspond to typical spatial scales of the involved processes: the low regime correspond to county level characteristics and the middle one the state level processes. This behavior confirms that prices are non-stationary in space, and that therefore appropriate statistical techniques must be used to study potential drivers at different levels. The two next subsections follow this idea and investigate potential independent variable to explain local fuel prices, using two different techniques corresponding to two complementary paradigms: geographically weighted regression that puts the emphasis on neighborhood effects, and multi-level regression taking into account administrative boundaries.

\subsection{Geographically Weighted Regression}

The issue of spatial non-stationarity of geographical processes has always been a source of biased aggregated analyses or misinterpretations when applying general conclusions to local cases. To take it into account in statistical models, numerous techniques have been developed, among which the simple but very elegant Geographically Weighted Regression (GWR), that estimates non-stationary regressions by weighting observations in space closely following kernel estimation methods. This was introduced in a seminal paper by~\cite{brunsdon1996geographically} and subsequently used and matured by many other studies. The significant advantage of this technique is that an optimal spatial range (optimal in the sense of model performance) can be inferred to derive a model that yields the effect of variables across space, revealing local effects that can occur at different spatial scales or across boundaries.

We proceed to multi-modeling to find the best model and associated kernel and spatial range. More specifically, we do the following: (i) we generate all possible linear models from five basic potential variables (which are income, population, average wage, jobs per capita, jobs); (ii) for each model and each candidate kernel shape (exponential, gaussian, bisquare, step), we determine the optimal bandwidth in the sense of both cross-validation and corrected Akaike Information Criterion (AIC) which quantifies information included in the model; (iii) we fit the models with this bandwidth. We choose the model with the best overall AIC, namely $price = \beta\cdot\left( income, wage, percapjobs\right)$ for a bandwidth of 22 neighbors and a gaussian kernel,\footnote{Note that the kernel shape does not have much influence as soon as gradually decaying functions are used} with an AIC of $2,900$. The median AIC difference with all other models tested is 122. The global R-squared is 0.27, which is relatively good compared to the maximum R-squared of 0.29.\footnote{Which was obtained for the model with all variables, which clearly overfits with an AIC of 3010; furthermore, effective dimension is less than 5 as 90\% of variance is explained by the three first principal components for the normalized variables.}

\begin{figure}
\centering
\includegraphics[width=\linewidth]{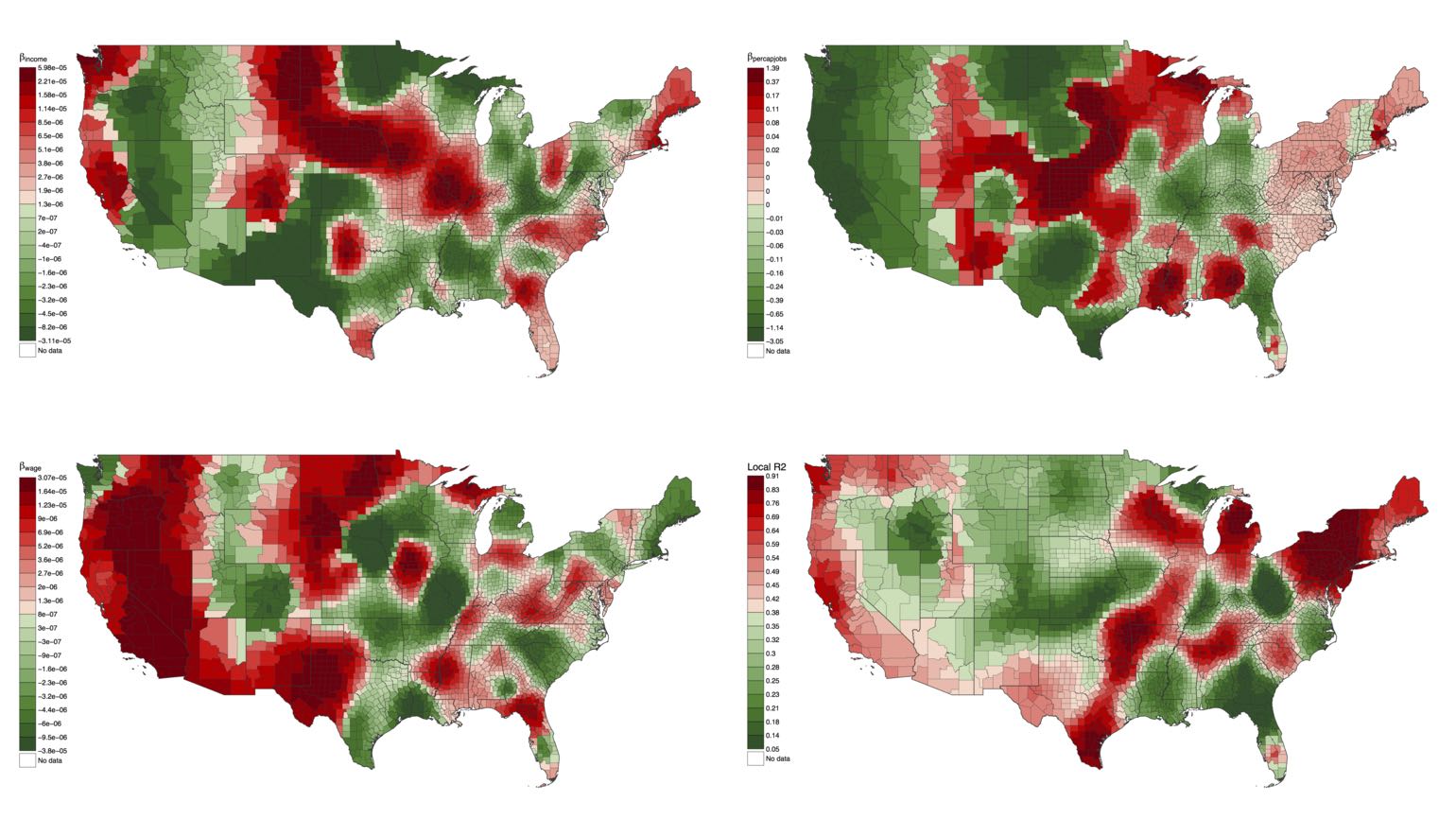}

\caption{\textbf{Results of GWR analyses.} For the best model in the sense of AICc, we map the spatial distribution of fitted coefficient, in order from left to right and top to bottom, $\beta_{income}$, $\beta_{percapjobs}$, $\beta_{wage}$, and finally the local r-squared values.}
\label{fig:gwr}
\end{figure}

The coefficients and local R-squared for the best model are shown in Fig.~\ref{fig:gwr}. The spatial distribution of residuals (not shown here) seems globally randomly distributed, which confirms in a way the consistency of the approach. Indeed, had a distinguishable geographical structure been found in the residuals, it would have meant that the geographical model or the variable considered had failed to translate spatial structure.

Let now turn to an interpretation of the spatial structures we obtain. First of all, the spatial distribution of the model performance reveals that regions where these simple socioeconomic factors do a good job in explaining prices are mostly located on the West Coast, the South Border, the North-East region from the Lakes to the East Coast, and a stripe from Chicago to the south of Texas. The corresponding coefficients have different behaviors across the areas, suggesting different regimes.\footnote{We comment their behavior in areas where the model has a minimal performance, that we fix arbitrarily as a local R-squared of 0.5} For example, the influence of income in each region seems to be inverted when the distance to the coast increases (from the North to South-East in the west, and from the south to the north in Texas), which could be a fingerprint of different economic specializations. On the contrary, the regime shifts for wage show a clear cut between west (except in the region of Seattle) and middle/east, that cannot be only attributed to state specific policies as Texas is clearly split in two. Accordingly, the influence of the employment rate shows an opposition between east and west, that could be attributed to cultural differences. These results are difficult to interpret directly, and must be understood as a confirmation that geographical particularities matters, as regions differ in regimes for each of the simple socioeconomic-variables. Further precise knowledge could be obtained through targeted geographical studies including qualitative field studies and quantitative analyses, that are beyond the scope of this exploratory paper and left for further research.

Finally, we extract the spatial scale of the studied processes, that is, we compute the distribution of the distance to nearest neighbors with the optimal bandwidth. The result yields roughly a log-normal distribution, with a median of 77km and an interquartile of 30km. We interpret this scale as the spatial stationarity scale of price processes in relation with economic agents, which can also be understood as a range of coherent market competition between gas stations.

\section{Econometric analysis}\label{sec:econometry}

In this section, we turn to a different analysis. Namely, we use multi-level linear regressions to analyse the extent to which some specific covariates explain the variation in the price of oil $\emph{within-state}$.

\subsection{Multi-level Regression}

Since our initial database allows us to look at the level of variable $x_{i,s,c,t}$, the fuel price in day $t$, in gas station $i$, in state $s$ and in county $c$, we start by running high dimensional fixed effect regressions following the model:
\begin{eqnarray}
x_{i,s,c,t} &=& \beta_s + \varepsilon_{i,s,c,t} \\
x_{i,s,c,t} &=& \beta_c + \varepsilon_{i,s,c,t} \\
x_{i,s,c,t} &=& \beta_i + \varepsilon_{i,s,c,t}
\end{eqnarray}

where $\varepsilon_{i,s,c,t}$ contains an idiosyncratic error and a day fixed effect. This first analysis confirm that most of the variance can be explained by a state fixed effect and that integrating more accurate levels has only small effect on the fit of our model as measured by the R-squared.

We now turn to a different reduced-form analysis, which aims at capturing the explanatory variables that account for spatial price variation of fuel. We consider the following linear model:

\begin{equation}
\label{eq:reg}
log(x_{i}) = \beta_0 + X_{i}\beta_1 + \beta_{s(i)} + \varepsilon_{i},
\end{equation}

where $x_{i}$ denotes average measured fuel price in county $i$ aggregated across all days, $X_{i}$ is a set of county specific variables. We chose these variables so as to cover different socio-economic local specificities of a county. The list of variable is given in Table \ref{tab:county_var_desc} and their descriptive statistics in Table \ref{tab:county_var_stat_desc}. In particular, we want to capture the potential use of cars by including density of the county and the proportion of resident using a car to commute. We also want to capture specific preferences by including the share of voters that chose a republican candidate during previous general elections. We complete these covariates by including a set of economic characteristics, controlling for the unemployment rate, the share of population with a bachelor degree, the poverty rate and the share of manufacturing sector. To complete the description of equation \ref{eq:reg}, we note $s(i)$ the state to which the county belongs so that $\beta_{s(i)}$ capture all state specific variation. Note that this fixed effect term removes taxes and that other coefficients are equivalent to an estimation on the orthogonal complement of the linear combination of states fixed effects, in particular they are estimated on before-taxes prices. Finally $\varepsilon_{i}$ is an error term satisfying $Cov(\varepsilon_{i}, \varepsilon_{j}) = 0$ if $s(i) \neq s(j)$. This clustering of standard error at the state level is motivated by findings of the previous section, showing that spatial autocorrelation of fuel price at the state level is still potentially strong. This specification aims at capturing the effect of various socio-economic variables at the county level after a state fixed effect has been removed.

We present our results in Table~\ref{tab:reg} where we progressively add more local covariates. Column (1) shows that regressing the log of price on a state fixed-effect and the log of the density already accounts for more than 76\% of the variance in price. This is mostly due to tax on fuel which are set at the state level in the US. In fact, when we regress the log of oil price on the level of state tax, we find a R-squared of 0.33\%. The fact that density is negatively associated with price can be explained by a simple competition model as presented in next subsection. One other explanation would be that gas station located in remote area must charge a higher price in order to cover their overhead costs.

The remaining explanatory variables have the expected signs: an more intensive use of cars is associated with lower prices just like more competition in the gas market (as captured by the number of gas stations per capita in the county) also pushes prices downward. The magnitudes of coefficients suggest that increasing the number of gas station per capita by 1\% decreases the price by around 0.6\%. That more competition is associated with a lower price is a basic result of the Industrial Organization literature and will be formally motivated in next subsection. 

Political preferences also matter, since gas price decreases with the extent to which a county has voted for a Republican candidate. This last finding suggests a circular link: counties that use car the most tend to vote to politician that promote pro car policies. This is discussed in \citet{hammar2004political} and is seen as one major obstacle to the setting of an adequate gasoline tax (see also \citealp{parry2005does}). \citet{Zitzewitz2013} shows that the price of fuel is used as a political tool, especially in swing states, by analyzing fluctuations just before important elections. It is interesting to note that the correlation between political preferences and gas price hold \emph{after} controlling for all other factors, including the fact that republican counties are less dense and more car intensive on average.

Finally, fuel price tends to be lower in poorer and less educated areas which we measure by an estimate of the poverty rate and the importance of the manufacturing sector, both of which being negatively correlated with price. The magnitude on the coefficient on poverty rate, combined with descriptive statistics in Table \ref{tab:county_var_stat_desc} suggests that moving from the first to the third quartile in terms of poverty is associated with a decline in oil price of around 0.5\%. 

Adding these explanatory variables slightly increase the R-squared, suggesting that even after having removed a state fixed-effect, the price of fuel can be explained by local socio-economic features.Among all these control variables, political preferences appear to be the best predictor of the residual local price of gas, after a state fixed effect has been removed.

\begin{table}[]
    \centering
    \caption{Source and description of the county level variables}
    \label{tab:county_var_desc}
    \begin{tabular}{lp{7cm}l}
    \hline
    Variable     &  Description & Source \\
    \cmidrule(r){1-1}
     \cmidrule(r){2-2}
      \cmidrule(r){3-3}
    \cr
    Density     &  Household units per sq miles & Census 2010 \\
    Unemployment & Total employed over labor force in 2017 & BLS \\
    Nb Stations & Number of gas station & Own calculation \\
    Manuf & Employment share of the manufacturing sector in 2016 & CBP - Census \\
    Bachelor & Share of population over 25 with a bachelor degree in 2017 & ACS \\
    Cars & Share of people over 16 using their car to commute in 2017 & ACS \\
    Poverty & Share of people considered in poverty by SAIPE program & SAIPE - Census \\
    Vote GOP & Share of voters that voted for a republican candidate in the general elections from 2000 & MIT election lab \\
    \hline
    \end{tabular}
\end{table}

\begin{table}[]
    \centering
    \caption{Descriptive statistics}
    \label{tab:county_var_stat_desc}
    \begin{tabular}{l ccccccc l}
    \hline
    Variable     &  mean & sd & p10 & p25 & p50 & p75 & p90 & units\\
    \cmidrule(r){1-1}
    \cmidrule(r){2-8}
    \cmidrule(r){9-9}

    \cr
    Density     &  113.4 & 823.8 & 2.2 & 8.4 & 21.0 & 51.5 & 164.2 & Households per sq mile\\
    Unemployment (x100)& 4.6 & 1.6 & 2.9 & 3.5&4.3 & 5.3 & 6.5 & Share\\
    Nb Stations & 28.0&65.9&1.7&3.3&9.4&24.7&64.9& Number \\
    Manuf (x100) & 19.3&17.0&0&6.1&15.4&28.5&43.0& Share \\
    Bachelor (x100) & 21.2&9.3&12.1&14.7&19&25.3&33.7&Share \\
    Cars (x100) & 89.5&7.3&82.5&87.8&91.3&93.5&95 &Share\\
    Poverty (x100) & 15.4 & 6.2 & 8.7&10.9&14.4&18.4&23.4 &Share \\
    Vote GOP (x100) & 59.4&13.2&41.8&51.4&60.5&68.7&75.5& Share\\
    \hline
    \end{tabular}
\end{table}

\begin{table}[h]
\caption{Ordinary Least Square estimations of Equation \ref{eq:reg}. Each model includes a state fixed effect. Robust standard errors are reported in parenthesis. ***, ** and * respectively indicate 0.01, 0.05 and 0.1 levels of significance.\label{tab:reg}}

\begin{tabular*}{\hsize}{@{\extracolsep{\fill}}lllllll@{}}
\toprule
  & (1) & (2) & (3) & (4) & (5) & (6) \\ 
\midrule
Density (log)&      -0.006***&      -0.006***&      -0.007***&      -0.008***&      -0.009***&      -0.011***\\
            &     (0.001)   &     (0.001)   &     (0.001)   &     (0.001)   &     (0.001)   &     (0.001)   \\

Cars       &               &      -0.151***&      -0.150***&      -0.187***&      -0.179***&      -0.165***\\
            &               &     (0.051)   &     (0.051)   &     (0.019)   &     (0.020)   &     (0.020)   \\

Nb Stations (log) &               &               &      -0.006***&      -0.005***&      -0.006***&      -0.006***\\
            &               &               &     (0.002)   &     (0.001)   &     (0.001)   &     (0.001)   \\

Vote       &               &               &               &      -0.040***&      -0.052***&      -0.041***\\
            &               &               &               &     (0.006)   &     (0.008)   &     (0.008)   \\

Poverty      &               &               &               &               &      -0.075***&      -0.063***\\
            &               &               &               &               &     (0.016)   &     (0.018)   \\

Manuf      &               &               &               &               &      -0.020***&      -0.014***\\
            &               &               &               &               &     (0.004)   &     (0.004)   \\

Bachelor       &               &               &               &               &               &       0.060***\\
            &               &               &               &               &               &     (0.013)   \\

Unemployment   &               &               &               &               &               &       0.238***\\
            &               &               &               &               &               &     (0.063)   \\

\cr
\cmidrule{2-7}
&       0.767   &       0.777   &       0.779   &       0.816   &       0.825   &       0.828   \\
Observations&        3059   &        3059   &        3059   &        3046   &        2993   &        2993   \\

\bottomrule
\end{tabular*}
\end{table}

As already stated, the use of these variables is motivated by the idea that they proxy for socio-economic and cultural local characteristics. We do not model the direct mechanism through which they affect gas price (nor did we in the case of GWR). However, to gain insight on such a mechanism, we now derive a simple model of monopolistic competition which can explain how the heterogeneity of the density of population can affect the price of gasoline.

\subsection{A minimal theoretical model} \label{sec:theory}

We introduce a minimal model to show how some of the above empirical stylized facts are consistent with a theoretical basis. In particular, the assumption that prices arise partly through the interactions between agents can be checked through a model including population density only. We adapt a spatial version of the Salop model introduced by \cite{salop1977bargains}. In this famous model, firms are located around a circle at regular intervals and consumers are uniformly distributed around the same circle. The firms can set the price but they take into account that consumers have to pay transportation cost. This results in an equilibrium price that is the same across all firms. Here we make two adjustments to this model in order to generate spatial heterogeneity. Namely we include (i) the fact that consumers are distributed according to a (known) population density and (ii) that the cost of transportation is itself a function of the firms' output (since in our case, firms are gas stations).\footnote{This model is of course very stylized, and in particular space is considered one dimensional (as an angle coordinate on a circle) to avoid edge effects.}

We now formally describe the model. There are N consumers indexed by $n=\{1..N\}$ and J gas station indexed by $j=\{1..J\}$. There are both located in a unit circle and we denote by $\theta(n)$ the position of consumer $n$ and $\phi(j)$ the location of station $j$. We assume that stations are located regularly so that $\phi(j) = \frac{2\pi j }{J}$ while $\theta(n)$ follows a known distribution $\mathcal{H}$ with density $h$.
Each station charges a price $p(j)$ and each consumer considers all possible options to buy gas and choose the cheapest. To purchase gas, a consumer needs to move to the gas station $j$ from its position at a cost $C(n,j) = \eta p(j) \left|\theta(n)-\phi(j) \right|$, where $\eta$ is a constant that captures the consumption of gas per unit of distance.

Let us consider a consumer $n$ located between stations $j$ and $j+1$ (formally $\frac{2\pi j }{J}<\theta(n)<\frac{2\pi (j+1) }{J}$). This consumer is indifferent between the two stations if $p(j)+C(n,j)=p(j+1)+C(n,j+1)$, that is if:

$$
p(j)\left(1+\eta\left(\theta(n)  -  \frac{2\pi j }{J}\right)\right) = p(j+1)\left(1-\eta\left(\theta(n)  -  \frac{2\pi (j+1) }{J}\right)\right)
$$

This yields:

$$
\theta(n) = \frac{1}{\eta}\frac{p(j+1)-p(j)}{p(j+1)+p(j)} + \frac{2\pi j}{J}+\frac{2\pi}{J}\frac{p(j+1)}{p(j+1)+p(j)} \equiv \theta(j, j+1)
$$
the position of the indifferent consumer. Hence, the profit of gas station $j$ can be written:

$$
\pi(j, p(j), p(j+1), p(j-1)) = \left(p(j)-P\right) \int_{\theta(j-1, j)}^{\theta(j, j+1)}{ d\mathcal{H}(\theta)}
$$

Where $P$ is the unit cost of the gas station input that we assume to be constant (we think at it as the price of crude oil). Each gas station $j$ chooses the price $p(j)$ that maximizes its profit, taken as given the price vector of all other gas stations.

First order condition (FOC) from this maximization problem yields:
$$
\int_{\theta(j-1, j)}^{\theta(j, j+1)}{ d\mathcal{H}(\theta)}+(p(j)-P)\left[\frac{d\theta(j,j+1)}{dp(j)}h(\theta(j,j+1)) - \frac{d\theta(j-1,j)}{dp(j)}h(\theta(j-1,j))\right] = 0,
$$
with 
$$\frac{d\theta(j,j+1)}{d(p(j))}=\frac{-p(j+1)}{\left(p(j)+p(j+1)\right)^2} \left(\frac{2}{\eta}+\frac{2\pi}{J} \right)<0$$
and
$$\frac{d\theta(j-1,j)}{d(p(j))}=\frac{p(j-1)}{\left(p(j-1)+p(j)\right)^2}\left(\frac{2}{\eta} +\frac{2\pi}{J}\right) > 0$$

Finally (FOC) can be rewritten:

\begin{equation}
\label{eq:tosolve}
\int_{\theta(j-1, j)}^{\theta(j, j+1)}{ d\mathcal{H}(\theta)} = \left(\frac{2}{\eta} +\frac{2\pi}{J}\right)(p(j)-P)\left(\frac{p(j+1)h(\theta(j,j+1))}{\left(p(j)+p(j+1)\right)^2} +  \frac{p(j-1)h(\theta(j-1,j))}{\left(p(j)+p(j-1)\right)^2}\right)
\end{equation}

We solve this system of $J$ equations to derive the distribution of equilibrium prices $\vec{p} = p(j)$. The model does not allow for a tractable solution for any distribution $h(\theta)$. We therefore adopt a numerical optimization strategy to obtain the price vector. We rewrite equation~\ref{eq:tosolve} as a minimization problem $f_j(\vec{p},h,J,\eta,P) = 0$ with the single objective $C\left[h,J,\eta,P\right] = \sum_j f_j(\vec{p},h,J,\eta,P)^2$ (we take $C$ as a standard squared cost function) that can be solved conditional on the parameters with an optimization heuristic. We rely to a genetic algorithm to solve the problem, as this type of heuristic is very flexible in the fitness landscapes handled and in the input space dimensionality. In particular, we use the R package GA (see \citealp{Scrucca2013ga}), with default setting, a population of size 50 and a maximal number of iterations of 10000. Each optimization is repeated 50 times following a Monte Carlo method for a better consistency of final results as the optimization is stochastic and does not always fully converge.

\begin{figure}
\centering
\label{fig:thmodel}
\subfigure[Uniform]{\label{fig:thmodel_uniform}
	\includegraphics[width=0.3\linewidth]{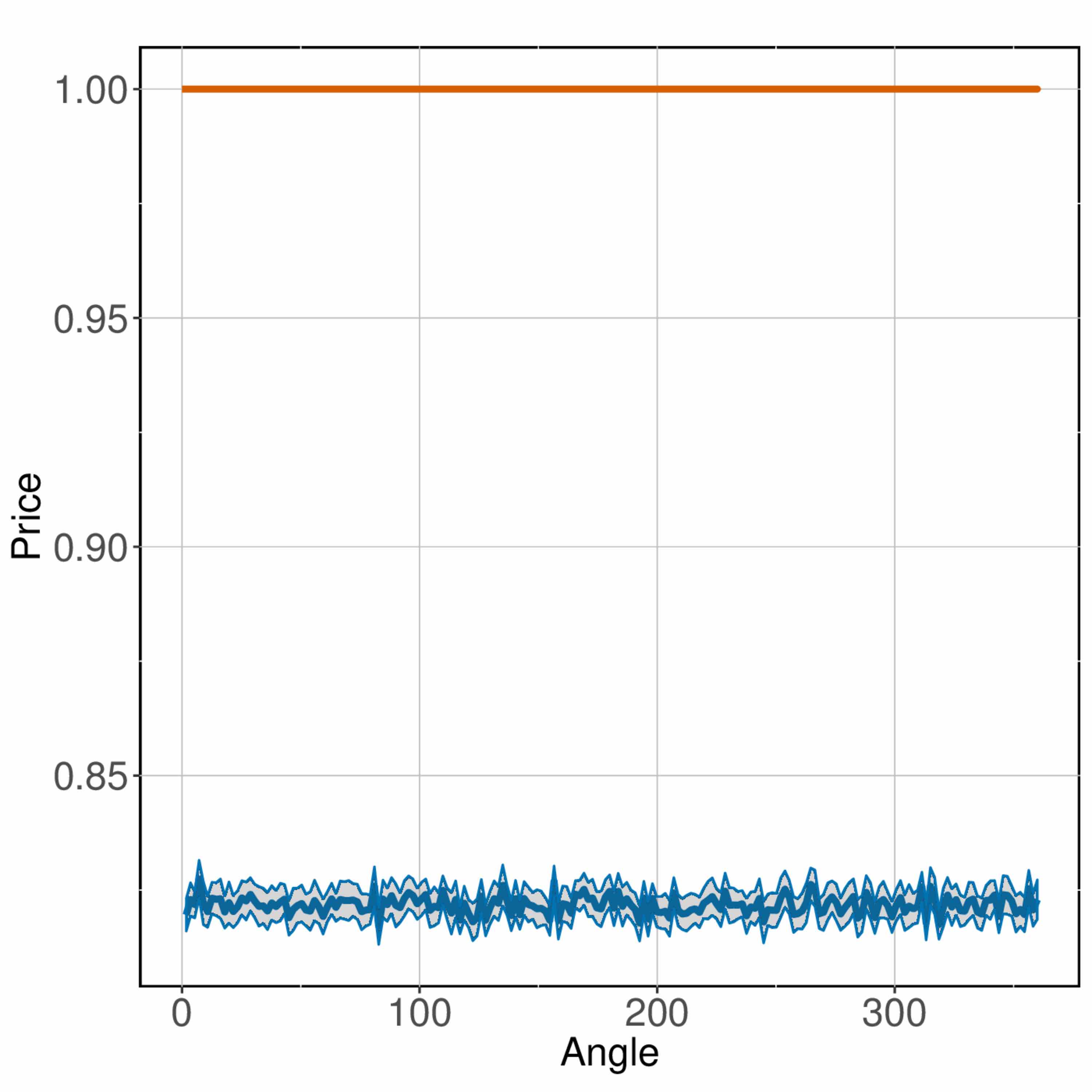}
    }
\subfigure[Linear]{\label{fig:thmodel_linear}
	\includegraphics[width=0.3\linewidth]{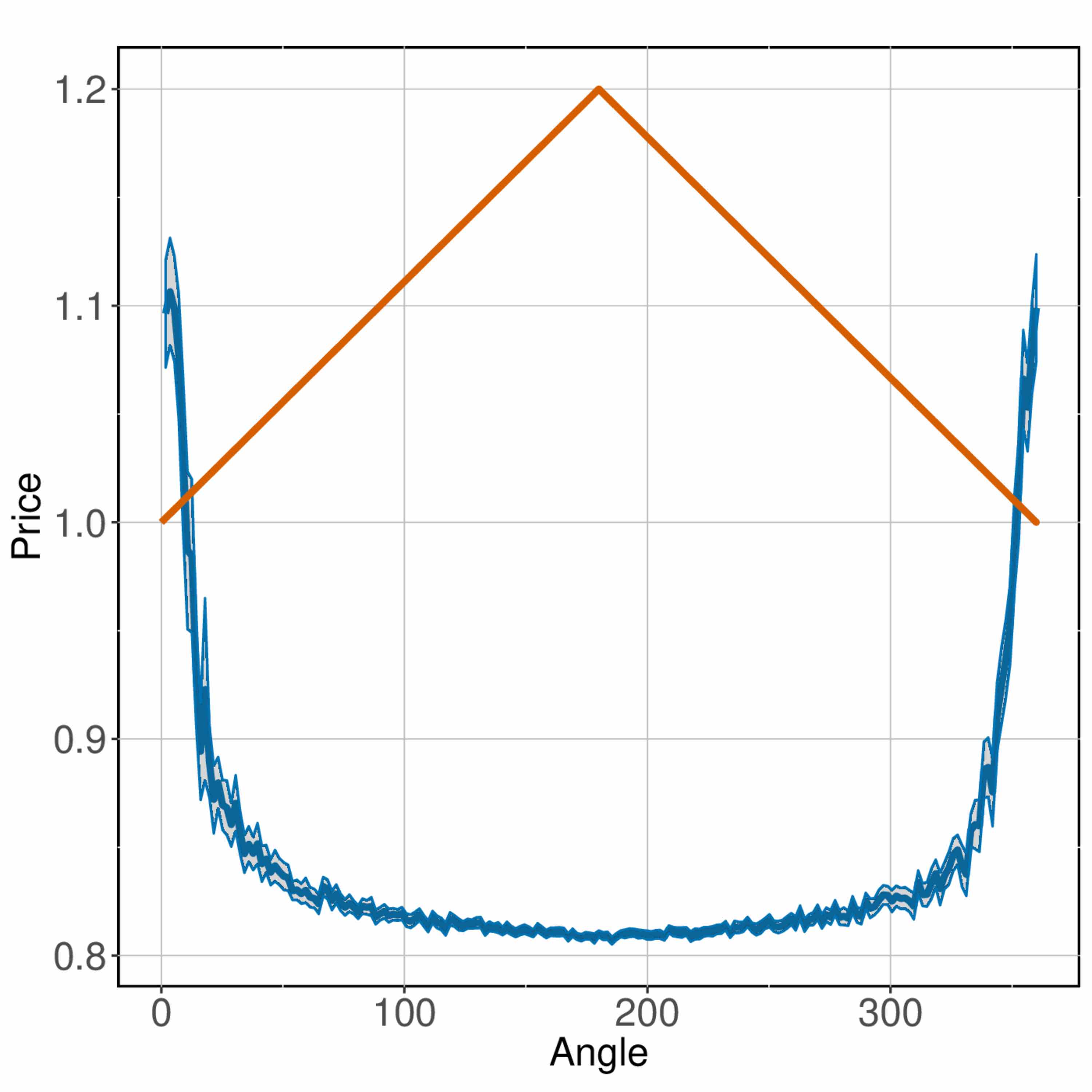}     
}
\subfigure[Exponential]{\label{fig:thmodel_expo}
	\includegraphics[width=0.3\linewidth]{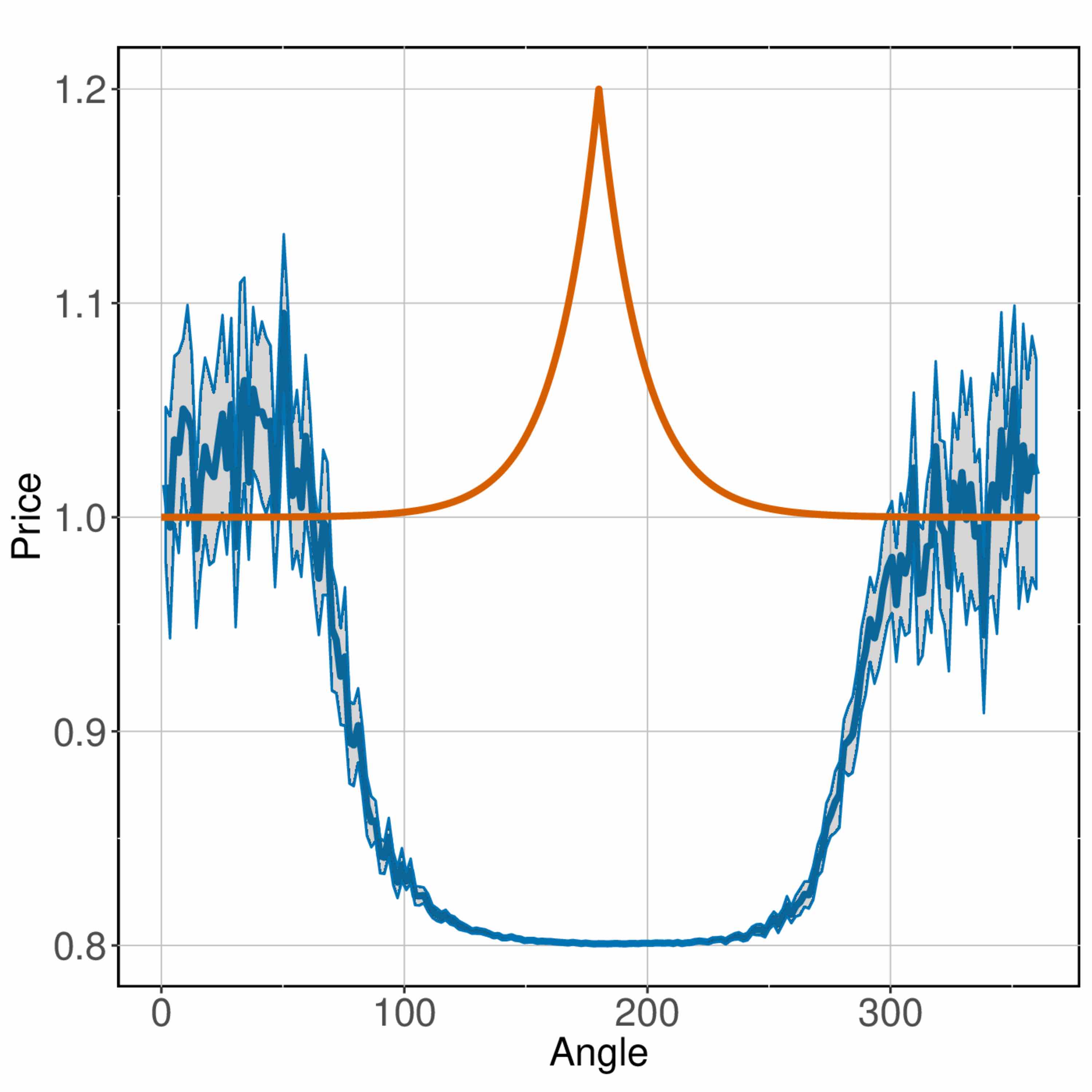}     
}
	\caption{\textbf{Numerical resolution of the theoretical model.} We give for each stylized population density profile (columns) within uniform, linear and exponential, the equilibrium prices as a function of the angle $\theta$ (here in degrees), estimated with the genetic algorithm optimization heuristic. The number of station is set to 200. Points are averages on the repetitions of the optimization, and shaded areas give corresponding confidence intervals.}
\end{figure}

Results from this numerical simulation are shown in Fig.~\ref{fig:thmodel} with fixed transportation cost $\eta = 1$, unit price $P=0.8$, $J=200$ gas stations and three different population density profiles: (i) a uniform population with $h(\theta) = 1 / 2\pi$ (left-hand side panel); (ii) a linear population such that $h(0)=0$, $h(\pi)=1 / \pi$ and $h$ varying linearly between the two (central panel); (iii) a shifted exponential population such that $h(\theta)=\exp ( - \left| \theta - \pi \right| / \theta_0)$ with $\theta_0 = \pi / 10$ such that most of the population is located between $\pi / 2$ and $3 \pi / 2$ (right hand side panel). This last profile is typical of urban population distributions as argued in \citet{anas1998urban}.

As expected, the uniform density profile in Figure \ref{fig:thmodel_uniform} yields uniform prices.\footnote{The perturbation comes from a small noise that results from the stochasticity of the optimization procedure, which is furthermore higher when the problem becomes more complex, i.e. when the number of stations increases. This noise does not affect the qualitative conclusions of the simulation.} The price profiles for the linear density in Figure \ref{fig:thmodel_linear} show a slight increase around the maximal population point, and a sharp increase when the population approaches the minimum. This effect is more pronounced with a higher number of stations in which cases areas with a very low population density end up with significantly higher prices. This trend is confirmed by the exponential population profile \ref{fig:thmodel_expo}, where prices rises as soon as a certain population threshold is attained. These qualitative results are robust to the number of stations $J$, as similar behaviors were obtained with $J\in \{10;20;100\}$.

Although extremely stylized, the model succeed in predicting our main findings which are that gas price are heterogneous in space and that they are negatively correlated with local population density.\footnote{In this model, we have used population and density interchangeably. In practice, the underlying link between the variation of density and the variation of population of a county is complex and out of the scope of this stylized model. Such relationship typically relates to more complex geographical processes (links between surface, urban density and population) which are not included here. These can for example be tackled at the macroscopic scale with urban scaling laws \citep{marshall2007urban} or at the mesoscopic scale with metropolitan models of population distribution \citep{anas1998urban}.} Here our main channel is through competition: for a given number of gas stations, a more densely populated area correspond to higher profit, the equilibrium price of a given station results in a trade off between decreasing its price in order to ``steal'' consumers from its neighboring competitors or increasing its price so as to raise its markup. The former strategy is all the more efficient that the population around the station is large.

While the main goal of the model is to show how more densely populated areas are associated \emph{ceteris paribus} with lower gas price, it can be easily related to other covariates that are analyzed in previous section. For example, the fixed cost $P$ can be related to the local level of price, which in turns is influenced by local economic profile (captured by the poverty rate, share of manufacturing sector etc...). The model could also further be extended by considering more heterogeneity (i.e. adding an heterogeneity of income on top of the spatial heterogeneity) or by considering different preferences (as captured by the vote for the Republican party in Table \ref{tab:reg}). Another interesting extension would be to consider a discontinuity in the circle that would reflect a frontier between two different states with two different level of taxes. All these extensions are left for further research.

Finally, note that in this simple version of the model, the number of gas stations and their location is exogenous and constant. As shown in the description of the dataset above, this assumption is not absurd given the low correlation coefficient between population and number of stations at a local level in the most populated areas. This however leaves room for an improvement of the model, and in order to fully specify it, one would need to impose two additional features: (i) gas stations are allowed to enter as long as their expected profit is positive and (ii) a new entrant can choose its location. Qualitatively, such addition to the model would reinforce the competition effect and highlight the negative correlation between the density of gas stations and price. In its current form, the model has been designed to show how a given grid of suppliers results in price heterogeneity through other factors.

\section{Discussion} \label{sec:discuss}

\subsection*{On the complementarity of Econometric and Spatial Analysis methods}

One important aspect of our contribution is methodological. We show that to explore a new panel dataset, geographers and economists have different approaches, leading to similar generic conclusions but with different paths. Some studies have already combined GWR and multi-level regressions (e.g. \citealp{chen2012using}), or compared them in terms of model fit or robustness (\citealp{lee2009determinants}). We take here a multi-disciplinary point of view and combine approaches answering to different questions, GWR aiming at finding precise predictor variables and to measure the extent of spatial correlation, whereas econometric models explain with more accuracy the effect of factors at different levels (state, county) but take these geographical characteristics as exogenous. We claim that both are necessary to understand all dimensions of the studied phenomenon.

\subsection*{Designing localized car-regulation policies}

Another potential application of our analysis is to help designing better car-regulation policies. Environmental and health issues nowadays require a reasoned use of cars, in cities with the problem of air pollution but also overall to reduce carbon emissions. \cite{fullerton2002can} showed that a taxation of fuel and cars can be equivalent to a taxation on emissions. \cite{brand2013accelerating} highlight the role of incentives for the transition towards a low carbon transportation. However, such measures cannot be uniform across states or even counties for obvious reasons of territorial equity: areas with different socioeconomic characteristics or with different amenities should contribute regarding their capabilities and preferences. Understanding local prices dynamics and their drivers, for which our study is a preliminary step, may be a path to localized policies taking into account the socioeconomic configuration and include an equity criterion. To this extent, it is interesting to compare the US: a large territory with heterogeneous taxes with other countries where the level of tax on oil is uniform.

\subsection*{Going further}

We insist that such an analysis is only preliminary. Our goal is to derive simple insights of what a large scale micro dataset can teach about time and space variations of oil price. Of course the application of such dataset, providing it can be collected on a larger time interval are manifold. In addition to shedding new light on the literature already mentioned in the introduction (estimation of a pass through of tax to gasoline price, asymmetric incidence of fuel tax, effect of natural disasters etc...), such dataset could be used to calibrate theoretical models of consumer and producer behaviors.

One other interesting research direction would be to look at gas station located at the border and to see how they react to change in fuel tax rates in a neighboring states. Standard models of industrial organization suggest that they should also adjust their price and increase their markups as a result of a reduction in local competition. More generally, the spatial diffusion of such local changes in tax across other areas that are not directly affected is a subject that deserve a more careful look, which can be provided by the use of very disaggregated data. 

Other developments include similar analysis on a longer time span, developing spatio-temporal models that factor in the temporal dimension that will arguably be more relevant in that case. This would however again require an extension of the dataset in time. Similarly, a qualitative microscopic validation of some conclusions obtained here through fieldwork and for example interview of retailers and companies \citep{melaina2017investing}, would be an interesting development towards the coupling of quantitative and qualitative analyses.

Finally, a development that would exploit the full potential of the microscopic scale of our dataset would be the development and parametrization of a spatially-explicit agent-based model for the retail fuel market at the microscopic level, similarly to the study done by \cite{heppenstall2005hybrid}. Indeed, our simple model introduced above was solved through optimization. Yet, an alternative would be to simulate the full model with explicit agents and consumption processes. This would allow to generalize the model to a more realistic geographical setting (including real road network and stations positions), which we could augment with the actual price information from our dataset. This would allow to investigate much more different variables and processes including socio-economic aspects, and for example be a way to test the impact of territory-specific policies discussed above.

\section{Conclusion}

We have described a first exploratory study of US fuel prices in space and time, using a new database at the gas station level spanning two months. Our first result is to show the high spatial heterogeneity of price processes, using interactive data exploration and auto-correlation analyses. We proceed with two complementary studies of potential drivers: GWR unveils spatial structures and geographical particularities, and yields a characteristic scale of processes around 75km; multi-level regressions show that even though most of the variation can be explained by state level characteristics, and mostly by the level of the tax on fuel that is set by the state, there are still socioeconomic idiosyncrasies at the county level that can explain spatial variations of fuel price. The effect of density is consistent with a minimal theoretical model linking population distribution with the formation of prices. Our contribution paves the way for further detailed studies using similar micro-level datasets.

\end{document}